\documentclass[11pt]{article}
\usepackage{graphicx}
\usepackage{amsmath}
\usepackage{amssymb}
\usepackage{amsxtra}
\usepackage{adjustbox}

\title{A new view on cosmology, with non-translational
invariant Hamiltonian }
\author{H.B. Nielsen\footnote{Speaker at the  Work Shop
    ``What comes beyond the Standard Models'' in Bled.},
  Niels Bohr Institut,}
  
\begin{document}
\maketitle
\begin{center}
{\large  Masao Ninomiya\\ 
  %and Yasuhiro Sekino
Yukawa Institute for Theoretical Physics,
Kyoto University, Kyoto 606-0105, Japan
and
Yuji Sugawara Lab., Science and Engineering,
Department of Physics Sciences, Ritumeikan university
E-mail: msninomiya@gmail.com }
\end{center}

%\date{``Bled''   , July , 2022}

``Bled'' july 2022

\begin{abstract}
  The idea of this contribution is to suggest a way to get rid of gravity
  as a dynamical space time approximately in cosmology and thus be able to
  use Hamiltonian formulation ignoring the gravitational degrees of freedom,
  treating them just as background. Concretely we suggest to use a back
  groud De Sitter space time and then instead of the usual choice of
  coordinates leading to a picture in which the Universe Hubble expands, we
  propose to identify the time translation in the new coordinate system with
  a Killing form transformation for the De Sitter space time. This then leads
  to unwanted features like the descripton being formally not translational
  invariant, but we have in mind just to get in a simple way time translation
  and its associated Hamiltonian, and shall then in word give some ideas of the
  from this point of view way of looking at the usual cosmology.
  
\end{abstract}


% optionally

\section{Introduction}\label{s:intro}

In quantum mechanics and in analytical mechanics one works with
very general mechanical systems using a Hamiltonian formalism, in which
the time $t$ is taken as a parameter as a function of which then the state
$|\psi(t)>$
is considered. In relativity theory and especially in general relativity
the time concept is complicated by being at the end a general coordinate,
which one has to {\bf choose}, and it cannot be treated correctly unless one
includes the gravitational field degrees of freedom.

But if we have some ideas developped in analytical mechanics or quantum
regi with a simple Hamiltonian not including gravitational degrees
of freedom and would like at a first crude stage to apply it to cosmology, then
we would like to be allowed to have at least a crude cosmology, in which
the gravitational field is considered a static background, so that
most importantly an expansion of space can be ignored. Of course one
could alternatively introduce as a dynamical variable the size of the
Universe, $a$ say, but that is really beginning to approximate a
{\em dynamical} gravity, which it is the purpose of the present
idea to avoid.

Let us at least state, that we want  in the ``central region''
in 3-space to have a flat space approximation like one really in a short
distance perspective usually work with in the neighborhood of our Milky Way.
Then other usual requirements which may not be so important for making
a Hamiltonian description o.k., such as the translational invariance
or the associated asssumption, that crudely there is the same density of
galaxies etc. all over on a very large scale, we do not need, if it is
troublesome to obtain.

Let us take as a first approximation cosmological model the De Sitter space
time model. (You may actually choose between taking the cosmological constant
either the effective one in the inflation era or the present effective value.)

Let us resume and concretize our ``model'':

\begin{itemize}
\item We take a De Sitter space time.
\item We take the time development to be identified with a Killing
  transformation of the space time approximating the cosmology (i.e.
  a Killing transformation for the De Sitter space time.
\item We arrange the Milky Way to be, where the ``new'' time translation
  operation deviates the least from the ``usual''
  FLRW (Friedmann–Lemaître–Robertson–Walker metric)
  parametrization ``time''.
\end{itemize}

  \subsection{Why we Like Hamiltonian formalism, but Trouble with Gravity}

  \begin{itemize}
  \item From quantum mechanics we get (historically ?) accustomed to
    work with theories described by a Hamiltonian.
  \item In general relativity the for the Hamiltonian so basic concept,
    the energy $E$ becomes strongly gauge dependent in the for cosmology
    interesting situations.
  \item So it looks at first, that one needs a quantum gravity; but that is
    awfull, because many colleagues work on that without being even themselves
    convinced so much. May be string theory is good but not immediately
    usefull for cosmology?
    \end{itemize}
 % \end{frame}
%\begin{frame}
 \subsection{Our Suggestion: Use a Killing Symmetry for ``Time Translation'' in
    Approximate Cosmology}

  The main suggestion of the present work/talk is:
  \begin{itemize}
  \item Get rid of gravity by taking the gravitational field - the geometry -
    as only a background field. I.e. do not include gravity in the dynamical
    degrees of freedom being treated by the Hamiltonian.
    
  \item But then we need the time translation symmetry to be at least an
    approximate symmetry of the gravitational degrees of freedom.
    
  \item So choose an approximately cosmologically correct geometry and
    identify the ``time translation symmetry'' with a Killing transformation
    symmetry of the approximate geometry.
    
    \end{itemize}
  %\end{frame}
  
  \section{DeSitter space time}
  %\begin{frame}
  
\begin{figure}h
    
  \includegraphics[scale=0.7]{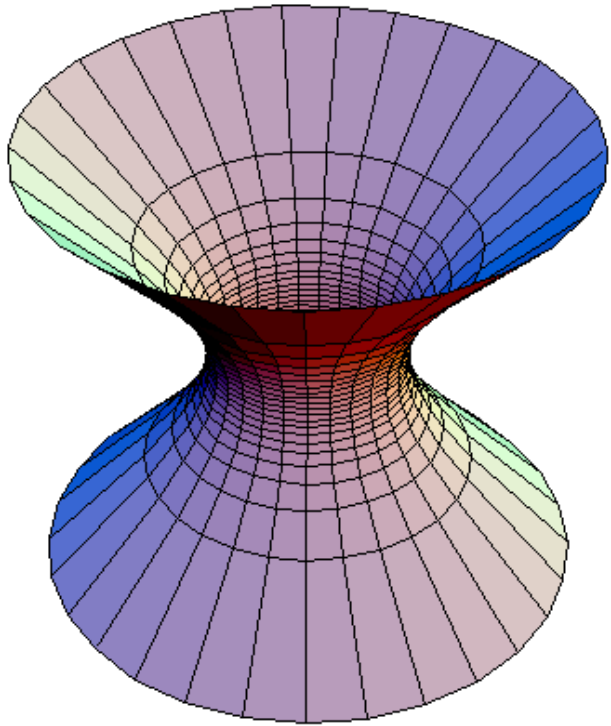}
  %\end{frame}
  \label{f1}
  \caption{The imbedded 1+1 DeSitter space time made to really represent
    the physically relevant 4+1 imbedding of the 3+1 dimensional
    De Sitter space time, which may be a crudely good cosmology.
    Here the coordinates drawn on the figure are the usual FLRW
    coordinates, in which the universe in the upper (= late time
    part) is expanding. The lower part will in most sensible
    models be considered so wrong that we should ignore it.}
  
\end{figure}

  To obtain a pictorial image of de Sitter space time we want to present
  a perspective picture in 3 dimensions to illustrate the imbedding of the
  3+1 dimensional De Sitter space into a 4+1 dimensional space-time, just
  for giving the illustration. But to do that we then need to simply remove
  2 of the spatial dimentions so  as to reduce 4+1 to 2+1 (correponding to
  what humans can conceive of as perpective drawing): See fig \ref{f1}.
  
 % {\bf De Sitter with Coordinates of Killing Symmetric type}
  
 % \includegraphics[scale=0.3]{DesitterMoschella.eps}

 %  {\bf De Sitter space with 1+1 dimension (instead of 3+1)}

 % {\bf De Sitter with Coordinates of Killing Symmetric type}
\begin{figure}  
  \includegraphics[scale=0.6]{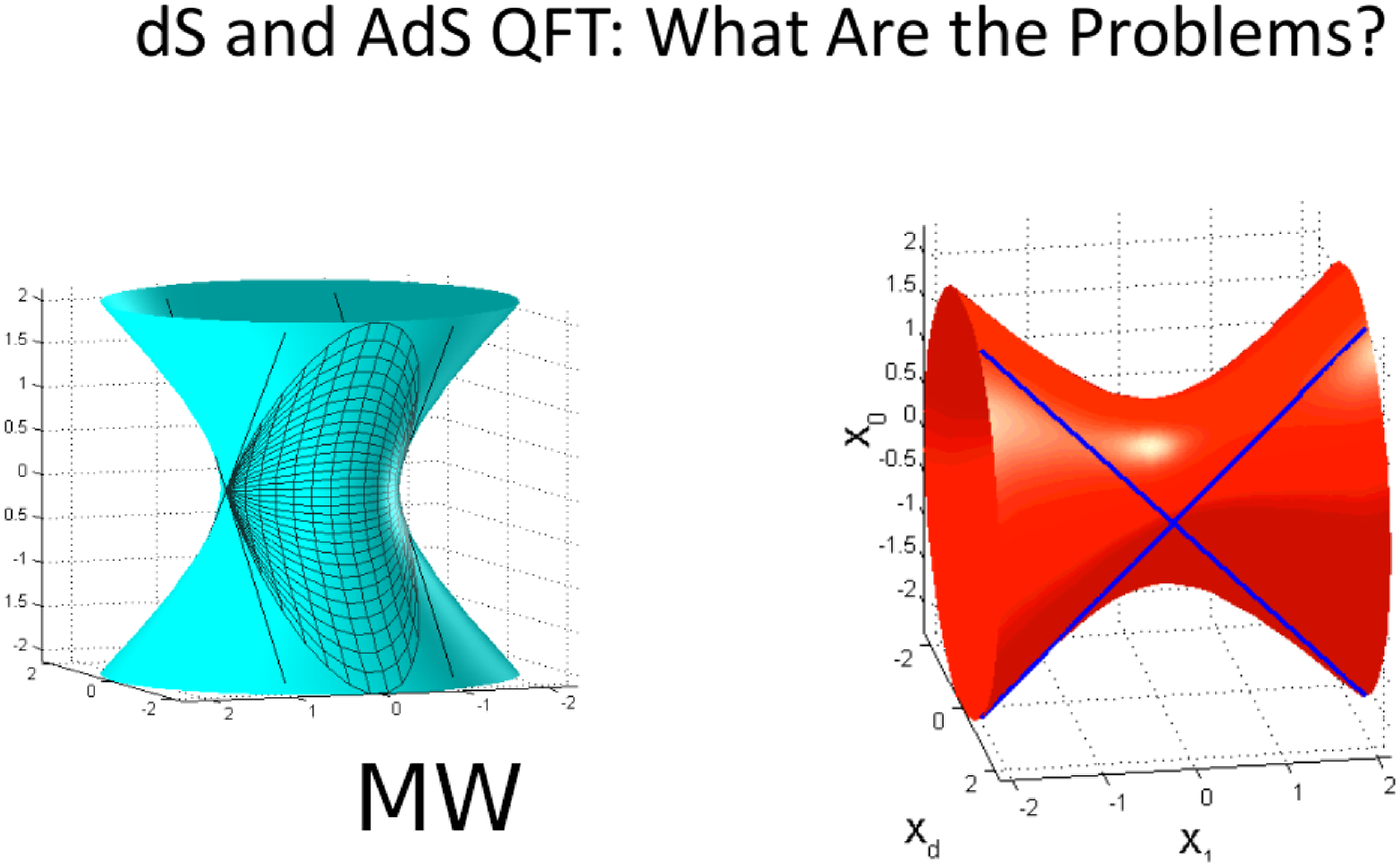}
 \label{f2}
\caption{In this figure we see both a de Sitter space time
    imbedding and an anti-Desitter space time imbedding.
    In both figures the time in the front region goes upwards on the
    figures. But we are in the present article really only
    interested in the De Sitter space-time in green to the left - and also do
    not care for the problem of what is wrong with quantum field theories -, and
    this figure illustrates {\bf De Sitter space with 1+1 dimension
      (instead of 3+1)}. On the one side of it is drawn a lap with
    coordinates, illustrating those coordinates we propose here: the
    coordinate curves going upward are the time coordinates and as
    such in the 1+1 dimension each represent a point in space. The more
    horizontal coordinate lines are ``parallel'' to the space
    coordinate and each of the lines represent a moment of time in the
    `` our '' coordinate system. The region below/earlier than the narrow
    neck is not to be taken seriously in usual cosmology, since it would be
    before the {\bf smoothed out} big bang.
}
\end{figure}
\begin{frame}

  \section{Coordinates}

  In the De Sitter model the space had at a certain time in the usual
  FLRW (fig.\ref{f1}) coodinates a most narrow i.e. least spatial
  size (radius $R$) moment
  of time. This is of course not true if one believes in a genuine Big Bang
  model, so it is only the time somewhat after that moment of the narrow space
  that should be taken approximately seriously. Also in the Killing form
  suggested coordinates as on the figure \ref{f2} the region below the narrow
  neck is of course presumably not to be taken seriosly. 

  Denoting the radius of the universe at the most narrow moment by $R$
  we can write in the imbedding coordinates the equation for the De Sitter
  space time surface as imbedded in the 4+1 dimensional space time, with
  the time -like  coordinate $X^0$ going upwards on the shown figures.
  Introducing of course an extra coordinate compared to usual 3+1
  space time, say $X^4$ we have the following equation for the imbedded
  surface to be identified
  with the universe space time:
  \begin{eqnarray}
    (X^1)^2+(X^2)^2 +(X^3)^2 +(X^4)^2 -(X^0)^2 &=& R^2.
  \end{eqnarray}

  We put the Milky Way at the maximal value of $X^4$ for a given value
  of $X^0$, i.e. indeed, for Milky Way:
  \begin{eqnarray}
    X^4 &=& \sqrt{R^2+(X^0)^2} \hbox{ for Milky Way }.
    \end{eqnarray}
  If we now want to keep to our wish to let the time at the Milky Way
  be the eigentime there, then we are forced to both in usual
  coordinates and in the ``new ''ones to have
  \begin{eqnarray}
    \sqrt{\frac{(dX^0)^2-(dX^4)^2}{dt^2}}&=&1 \hbox{ along the track of Milky
        Way}
  \end{eqnarray}
  This in fact leads to
  \begin{eqnarray}
    X^4(t)_{MW} &=& R * \cosh \frac{t}{R}\\
    X^0(t)_{MW}&=& R*\sinh \frac{t}{R}.\label{X0MW}
    \end{eqnarray}
  where $t$ stands for the time coordinate $t_u$ of the usual FLRW coordinates
  or in the ``new'' proposal $t_n$.
  In the ``usual'' FLRW model we keep the equation (\ref{X0MW}) to be valid
  not only for the Milky way, but all over. Usually one then defines the
  radius of the Universe at the time $t_n=t$ by the equation for a $S^3$-sphere
  representing the space at that time:
  \begin{eqnarray}
    (X^1)^2 +(X^2) + (X^3)^2 +(X^4)^2 &=&a^2\\
    \hbox{so that } a &=& R*\cosh\frac{t_u}{R}.
  \end{eqnarray}

  In the ``new'', here suggested coordinates, we rather let the ``momement of
  time'' cut straight back in ``usual'' time to the $S^2$-sphere given by
  \begin{eqnarray}
    \hbox{``Axis sphere'' } X^0 = X^4 &=&0\\
    \hbox{or } (X^1)^2+(X^2)^2 +(X^3)^2 &=& R^2.
    \end{eqnarray}
  That is to say, that
  \begin{eqnarray}
    \frac{X^0(t_n)}{X^4(t_n)} &=& \tanh \frac{t_n}{R}
    \hbox{ for ``new'' system.}\\
    \hbox{so that for $t_n$ fixed } \frac{dX^0}{dX^4} &=& \frac{t_n}{R}.
    \end{eqnarray}

  Let us also define a distance ${Dist to MW}_n$ from the Milky Way along
  the equal time
  $t_n$ in the ``new'' coordinate system out to a running point counted
  spacelike by
  \begin{eqnarray}
    d\hbox{Dist to MW}_n^2 &=& da^2+(dX^4)^2 - (dX^0)^2\\
    \hbox{as function of the angle } \theta_n &=& \arccos \frac{X^4}{X^4_{MW}}
    = \arccos \frac{X^4}{R*\cosh \frac{t_n}{R} }\\
    \hbox{giving } \hbox{Dist to MW}_n &=& R\theta_n.\\
    \hbox{In fact using for fixed $t_n$ that } (dX^4)^2-(dX^0)^2 &=&\frac{1}
         {\cosh^2\frac{t}{R}}*(dX^4)^2\\
         \hbox{and } a^2+(X^4)^2&=& (X^4_{MW})^2= R^2(\cosh \frac{t_n}{R})^2\\
         \hbox{one gets } X^4 &=& \cos( \theta_n) * R*\cosh\frac{t_n}{R}\\
         \hbox{and  }a&=& \sin( \theta_n)*R \cosh \frac{t_n}{R}\\
    \end{eqnarray}

  %Here I insert a bit from the possibly to be deleted part:
  
\subsection{How to consider the Killing transformation in the Imbedding}  
  If we consider how a fixed point in the ``space'' in the `new'
  coordinates move as function of the time $t_n$ we can use that it is
  rotated in the imbedding space time with its indefinite metric - the
  $X^0$ being a time coordinate - around the three-space given as
  $X^4=X^0 =0$. That is to say that the distance to this three-space is
  constant as long as an event is moved just by progressing the
  ``new'' time $t_t$. The rate of running of the local eigentime
  relative to the coordinate time in our ``new'' system is thus proportional
  to the (Lorentz) invariant distance from the three-space to the point.

  %{\bf Here starts what I consider the successfull attempt to calculate, while
  %  many other formulas shall probably be deleted at the end, because they
  %  did not come to the end:}
  \subsection{Developping coordinate transformation}
  
  In the ``new'' coordinates the Milky Way
  %interesting
  coordinates in the
  imbedding system are given as
  \begin{eqnarray}
    (X^4_{MW},X^0_{MW}) &=& (R*\cosh\frac{t_n}{T_H}, R\sinh\frac{t_n}{T_H})\\
    \hbox{and thus for running point } &&\nonumber\\
    (X^4,X^0) &=& ((\cos(\theta_n) * R*\cosh\frac{t_n}{T_H},
    \cos( \theta_n)* R\sinh\frac{t_n}{T_H})
  \end{eqnarray}

  Since in the usual FLRW we have $X^0= R \sinh\frac{t_u}{T_H}$ we can put
  up the equation
  \begin{eqnarray}
    \sinh\frac{t_u}{T_H} &=& \cos(\theta_n)\sinh\frac{t_n}{T_H}.\label{cr1}
  \end{eqnarray}
  (Here we have written $T_H$ for Hubble time, a constant parameter of
  dimension time. The simplest is to take $T_H = R$.)
  
  Except at the Milky Way where $t_u=t_n$ we have in the whole
  (half) space $t_u<t_n$ meaning that the events with which
  we today see as simultaneous with our time in the ``new''
  coordinates belong to the past in the usual FLRW scheme.

  In the usual scheme
  \begin{eqnarray}
    X^4 &=& R*\cos (\theta_u) *\cosh\frac{t_u}{T_H}\\
    \hbox{and so }  R*\cos (\theta_u) *\cosh\frac{t_u}{T_H}&=& \cos(\theta_n)*R*
    \cosh \frac{t_n}{T_H}.\label{cr2}
  \end{eqnarray}
  Since we already saw that $t_u<t_n$ almost anywhere in the positive
  $X^0$ and thus relevant region, we have also here
  \begin{eqnarray}
    \cos(\theta_u)&>&\cos(\theta_n)\\
    \hbox{and thus } \theta_u &<& \theta_n,
  \end{eqnarray}
  where the difference between the two angles though gets smaller in absolute
  value (but we shall see below not relatively)
  the smaller the $\theta$ angles, and goes to equality at the Milky Way
  at the $\theta$s being zero.

  Because the time development in the ``new'' scheme is given as a Killing
  transformation the spatial geometric structure in this ``new'' coordinate
  system is constant as a function of the time $t_n$, so that say the
  radius $\frac{\pi}{2}*R$ of the ``half''-space remains of
  this value at all times $t_n$, while the corresponding radius of the
  ``half''-space in the ``usual'' FLRW coorordinates grows with the
  time $t_u$ as
  \begin{eqnarray}
    \hbox{``half''-space radius}_u &=& \frac{\pi}{2}*R*\cosh\frac{t_u}{T_H}
    \nonumber \\
    \hbox{so that the logarithmic derivative }
    \frac{ d \hbox{``half''-space radius}_u}
         { \hbox{``half''-space radius}_u dt_u}&=&
         \frac{1}{T_H}*\tanh\frac{t_u}{T_H}\nonumber\\
         \hbox{so $T_H$ is Hubble time for } \tanh\frac{t_u}{T_H} & \approx
         & 1.\nonumber
    \end{eqnarray}

  \subsection{Only a lap is in both coordinates}

  As one may see from the figure \ref{f2} also it is not the whole usual
  De Sitter space which is described in the ``new'' coordinates, but
  rather only a lap, because
  %however
  late in the ``new'' system one looks
  the simultaneity surface in the ``new'' coordinates must still be a
  space-like surface, and thus seen from the usual system the point
  motion represented by the $t_n$ fixed to a value, must run with bigger than
  light speed. For late $t_n$ times however it goes very close to the
  speed of light, except near the Milky Way, where $t_n$ and $t_u$ are
  approximately equal. But this means that there is no way to get an event
  so close to $ (\theta_u=0, t_u =0)$ that it could be reached by a
  signal from  $ (\theta_u=0, t_u =0)$ represented in the ``new'' system.
  This limit means that in order, that an event can be represented
  in the ``new'' coordinateswe must have
  
  \begin{eqnarray}
    t_u &\le &R* (\frac{\pi}{2} -\theta_U)
    \hbox{ approximately for small $t_u$ }\\
    \hbox{or more exactly: }    \cos (\theta_u) *\cosh\frac{t_u}{T_H} &\ge&
    \sinh\frac{t_u}{T_H}\\
    or \;1\ge  \cos\theta_u &> &\tanh\frac{t_u}{T_H}\\
   1\ge \cos\frac{\hbox{Dist to MW}_u}{R\cosh\frac{t_u}{T_H}}&>&
    \tanh\frac{t_u}{T_H}\approx 1 \hbox{for large $t_u$}\nonumber\\
     % \hbox{ or } R*\cos \theta_u &>& R*\tanh \frac{t_u}{T_H},
     % \hbox{ which for
     %   small $t_u$ etc gives the same.}\\
     % \hbox{ or } \hbox{Dist to MW}_u = R*\cosh\frac{t_u}{T_H} *\cos \theta_u
     % &>& R*\sinh\frac{t_u}{T_H}\approx \frac{t_u R}{T_H}\\
  \end{eqnarray}

  For very late times, i.e. $t_u \rightarrow \infty$ we have
  \begin{eqnarray}
    \tanh\frac{t_u}{T_H} &\approx &1-2\exp(-2\frac{t_u}{T_H})\\
    \cos\frac{\hbox{Dist to MW}_u}{R\cosh\frac{t_u}{T_H}}&\approx &
    1-2*\left (\frac{\hbox{Dist to MW}_u^2}{R^2\exp(-2\frac{t_u}{T_H})} \right )\\
  \end{eqnarray}

  Inserting these approximations of late time into the inequlity yields
  \begin{eqnarray}
    \hbox{Dist to MW}_u &<& R \hbox{for large $t_u$.}
    \end{eqnarray}
  That is to say that for late times there is still a constant - of magnitude
  $R$ the most narrow size of the De Sitter Universe - 
  radius region around the Milky Way in which transition to the ``new''
  coordinates is possible. Regions in the usual coordinates further
  away than that cannot be transformed into the ``new'' coordinates.
  The angle $\theta_u$ describing this transformable region of
  course falls exponentially with time $t_u$, which is natural since the
  size of the unviverse grow exponentially.

  \subsection{Develloping formulae}

  By division of our coordinate relations (\ref{cr1}) and (\ref{cr2}) we
  obtain
  \begin{eqnarray}
    \cos \theta_u &=& \frac{\tanh\frac{t_u}{T_H}}{\tanh\frac{t_n}{T_H}}
    \label{cosu} 
  \end{eqnarray}

  We can also just write (\ref{cr1}) and (\ref{cr2})  as respectively
  \begin{eqnarray}
    \hbox{(\ref{cr1}) as }\cos \theta_n &=&
    \frac{\sinh \frac{t_u}{T_H}}{\sinh\frac{t_n}{T_H}}\label{cr1r}\\
    \hbox{and (\ref{cr2}): }\frac{\cos\theta_u}{\cos \theta_n}&=&
    \frac{\cosh \frac{t_n}{T_H}}{\cosh \frac{t_u}{T_H} }\label{cr2r}
    \end{eqnarray}

  %{\bf There must be a mistake in the calculations somewhere because
  %  from these formulae you could derive that $t_u =t_n$, but that is
  %  not true in the coordinate transformation we study.}
  
  \subsection{Near Milky way}

  For small $\theta_u$ and $\theta_n$, i.e. near the Milky way we can of
  course approximate
  \begin{eqnarray}
    \cos \theta &\approx& 1-\frac{1}{2}\theta^2
    \end{eqnarray}
  and if we are mainly interested in late times we may also
  use approximations like
  \begin{eqnarray}
    \cosh\frac{t}{T_H}\approx \sinh\frac{t}{T_H} &\approx &
    \frac{1}{2}*\exp(\frac{t}{T_H})\\
    \tanh\frac{t}{T_H} &\approx & 1- 2\exp (-2\frac{t}{T_H}).
    \end{eqnarray}
  Note that in the same approximation as the $\sinh$ and $\cosh$ ones
  given here the $\tanh$ would be exactly 1.

  In these approximations of late time and small angles we get
  \begin{eqnarray}
    \frac{1}{2} \theta_u^2 &\approx & 2\exp(-2\frac{t_u}{T_H})
    -2\exp(-2\frac{t_n}{T_H})\\
    \hbox{or }\theta_u &\approx&2 \sqrt{\exp(-2\frac{t_u}{T_H})-
      \exp(-2\frac{t_n}{T_H})}\\
    1-\frac{1}{2}\theta_n^2 &=& \exp(\frac{t_u-t_n}{T_H})\\
    \hbox{Taking ln: } -\frac{1}{2}\theta_n^2&=&  \frac{t_u-t_n}{T_H}\\
    \frac{\cos\theta_u}{\cos \theta_n}\approx 1-\frac{1}{2}(\theta_u^2-
    \theta_n^2)&=&
    \frac{\cosh \frac{t_n}{T_H}}{\cosh \frac{t_u}{T_H} }
    \approx \exp(\frac{t_n-t_u}{T_H})\\
    \hbox{taking ln: }-\frac{1}{2} (\theta_u^2-\theta_n^2) &=&
    \frac{t_n-t_u}{T_H}. 
    \end{eqnarray}
  We see that here 
  \begin{eqnarray}
    \theta_u & <<& \theta_n\\
    \hbox{Indeed :}&&\nonumber\\
    \theta_u &\sim& \sqrt{\exp(-2t/T_H)} \sim \exp(-t)\\
    \hbox{while } \theta_n &\sim & \frac{t_n -t_u}{T_H}
    (\hbox{much bigger})
    \end{eqnarray}
  Really in the region near the Milky Way the two times $t_u$ and $t_n$
  only deviate little compared to their approximately common zize the age of
  the universe at the time considered, which here was taken to be large.
  We should have in mind that distance between galaxies or galaxy clusters
  are roughly constant in the angular coordinate $\theta_u$ in the usual
  coordinates, so that the diminishing exponentially as $\propto \exp(-t/T_H)$
  of $\theta_u$ relative to the ``new'' $\theta_n$ in the region around the
  Milky Way means that the galaxies - to keep their fixed coordinates in
  $\theta_u$ - seen in $\theta_n$ moves away as with a Hubble expansion as
  if the ``new'' angular coordinate meant a genuine distance. Actually
  $\theta_n$ means a genuine distnace since we already noted that
  \begin{eqnarray}
    \hbox{Dist to MW}_n &=& \theta_n*R.
  \end{eqnarray}
  It is namely so in the beginning at $t_n =0$, but since the development
  in the ``new'' coordinates is of Killing form transformation type, the
  spatial structure and metric does not change under the $t_n$ time development.
  So not surprisingly we see the Hubble expansion in the distance as seen in
  the ``new'' coordinates. Of course {\em this} Hubble epansion can{\em not}
  continue at longer distances from the Milky Way, since the whole spatial
  universe in the ``new'' coordinate system is bounded of the size $R$.
  So the galaxies must seen in the ``new'' system somehow collect up in the
  large $\theta_n$ region. We can consider this large $\theta_n$ region as
  a kind of garbage place where galaxies expelled from the neighborhood
  of the Milky Way are thrown in, and get concentrated there. These regions
  also contribute relatively less to the Hamiltonian, so it is natural to
  consider them not so important and a kind of garbage place.

  \section{Hubble Expansion}

  Having in mind that in the ``new'' coordinates radius of the treated part
  of the universe remains $R$ = the narrowest size in the ``usual'' coordinates,
  there is no possibility for a true expanding universe as a whole.

  But of course translated back to the ``usual'' LFRW coordinates we assume that
  there is the usual type of Hubble expansion. So what happens in our
  ``new'' coordinate system?

  To orient ourselves let us start by estimating the rather easy to
  calculate relative velocity of the galaxies or pregalactic material
  in the region near the $\theta=\frac{\pi}{2}$ boundary. Here the usual time
  $ t_u$ never gets very big even for huge times $t_n$ in our new coordinate
  system.
  The ``new'' system locally in this boundary region moves with
  velocity corresponding to a ``hyperbolic angle'' $\frac{t}{R}$, meaning
  that the relative velocity is
  \begin{eqnarray}
    v&=& \tanh \frac{t}{R}\\
    \hbox{so that } \hbox{``Lorentz contraction factor''} = \gamma^{-1}
      &=& \sqrt{1-v^2}=\frac{1}{\cosh \frac{t}{R}}\\
      \hbox{To compare to } \frac{a}{R} = \cosh \frac{t}{R}.
  \end{eqnarray}
  That is to say: The Lorentz contraction - for the moment in the
  region where it is most easy to calculate it - is just of the size needed
  to Lorentz contract the Hubble expansion away and to put the universe as
  conceived in the ``usual'' LFRW coordinates of radius $a$ into the
  universe as seen at all times in our ``new'' scheme as having the
  radius only $R$ (remember  $R<a$ all the time except in the first moment).

  Very close to the Milky Way the velocity of the ``new'' versus the
  ``usual'' local frames goes through zero and thus here the relative frame
  velocity is small. Thus very close to the Milky Way the Lorentz contraction
  is also small.

  Rather
  the Hubble expansion means that the galaxies move away from the Milky Way
  more and more into the boundary region close to $\theta_n \approx
  \frac{\pi}{2} $. So the major part as seen in the ``new'' coordinates gets
  more and more empty/vacuum, only the Milky Way because of our choice
  stands back in the midle.

  {\bf Here a shorter attempt on Hubble Expansion in our system:}

  In the usual coordinates of course the galaxies are static in the sense
  of having e.g. staionary $\theta_u$ values ideally. The distance along
  space in the usual coordinates are:
  \begin{eqnarray}
    \hbox{Dist to MW}_u &=& \theta_u*R*\cosh\frac{t_u}{T_H}\\
    \hbox{Dist to MW}_n &=& \theta_n*R.
    \end{eqnarray}

  Thus in the usual frame the expansion of the distance from
  the Milky Way to a galaxy at $\theta_u$ goes as
  \begin{eqnarray}
    \hbox{Dist to MW}_u &=& \theta_u*R*\cosh\frac{t_u}{T_H}\\
    \hbox{and log-derivative }
    \frac{d\hbox{Dist to MW}_u}{\hbox{Dist to MW}_u dt} &=&
    \frac{1}{T_H}\tanh\frac{t_u}{T_H}\\
    &\approx& \frac{1}{T_H} \hbox{ for late times.}
    \end{eqnarray}
  
  \section{Time}

  It should be pointed out that the simulatneity in our ``new'' coordinates
  is closer to the way astronomers have to think about the happenings
  in practice: Because of the time it takes the light or also gravitational
  waves say to run we would be tempted to think of what in usual coordinates
  happened as long a go as the happening is light years away,as if it
  happened now. In our ``new'' coordinates the simultaneity sufaces
  are much closer to this tempting point of view. In our scheme there is
  a part of the big bang creation still present very far out near the
  bourder of $\theta_n=\frac{\pi}{2}$.

  %{\bf The remaining part here has presumably been replaced by the above
  %  additions:}
  %If we to avoid going into the for us here somewhat unimportant
  
  \section{The Non-translatioanl invariant Hamiltonian}
  It was the major disadvantage of our proposal that we should give
  up translational invariance along space. In fact the point is that
  the time progress in the ``new'' coordinate system to be conceived of as
  the development due to the Hamiltonian should represent a Killing form
  development corresponding in the imbedding we have used so much to a space
  time rotation (so it is really boosting). The crux of the matter now is
  the genuine time progress is thus much smaller in the region around which the
  rotation goes. Thus not unexpected the original Hamiltonian which would
  crudely to be used in the FLRW system should be deminished in this
  region where the timeprogressing is slow. In fact this diminishing goes
  proportional $\cos\theta_n$ or equivalently
  $\cos\frac{\hbox{Dist to MW}_n }{R}$. Thus the Hamiltonian to be used
    would rather be:
  
  {\bf To be Used Hamiltonian}

  \begin{eqnarray}
    H& =& \int \cos(\frac{``\hbox{Dist to MW}''_n}{R}) {\cal H}(x) d^3\Omega
  \end{eqnarray}
  where
  %$dist$
  $``\hbox{Dist to MW}''_n$is the distance to the Milky Way, $R$ the radius
  parameter in
  the De Sitter space used, ${\cal H}(x)$ the Hamiltonian density in a
  usual sense. Note that only half the De Sitter space is in the proposed
  region, or rather in the other half the usual energy is counted with a
  {\bf negative} weight factor!
  \end{frame}

\section{Conclusion}

We have proposed to use a De Sitter approximation as a back ground ansatz for
the gravitational fields metric tensor so that back reaction can, although
rather
approximately only be ignored, and thus gravity can be kept out of the
study provided one can get rid of such to non-gravitational theory
not usual effects as the Hubble expansion as a room expansion. The proposal
is to get rid of the Hubble expansion as an effect from space expansion
by choosing a coordinate system corresponding to the Killing transformation of
one of the symmetries of the De Sitter space-time to be identified with the
time progress transformation. The price of this choice is, that we loose
translational invariance in space, although gaining it in time to make up for
it (time translation
symmetry is of course violated if the varying size of the universe is
considered a back ground effect.) But for using the usual quantum mechanics
formalism with a conserved Hamiltonian a scheme as the one here with broken
translational invariance in space but unbroken in time is preferable.
Around the Milky Way we could arrange to have approximately the
special relativity in the coordinates of the ``new'' system.
The region most far away from this Milky Way has very strongly suppressed
Hamiltonian and thereby time development and remain at the satge of the
ealy universe almost forever. We suggested treating these far away from Milky
way places as kind of garbage place in which more and more of  Hubble expanded
material will end up, and since it contributes suppressed to the energy
it is not so bad to suggest to ignore these far away regions nearer
$\theta_n =\frac{\pi}{2}$.

%\end{document}

\section*{Acknowledgements}
It is a pleasure to thank Yasuhiro Sekino for discussions and especially
a Zoom-session about the Susskind article which use also such coordinates at
a point. Holger Bech Nielsen thanks the Niels Bohr Institute for status as
emeritus.

Masao Ninomiya acknowledges Yukawa Institute of Theoretical
Physics, Kyoto University, and also the Niels Bohr Institute and Niels Bohr
International
Academy for giving him very good hospitality during his stay. M.N. also ac-
knowledges at Yuji Sugawara Lab. Science and Engeneering, Department of
physics sciences Ritsumeikan University, Kusatsu Campus for allowing him as
a visiting Researcher

%% The bibliography section

\end{document}